# Superconducting properties and gap structure of the topological superconductor candidate Ti$_3$Sb

R. Chapai,[1,*] M. P. Smylie,[1,2] H. Hebbeker,[2] D. Y. Chung,[1] W.-K. Kwok,[1] J. F. Mitchell,[1] and U. Welp[1]

[1]*Materials Science Division, Argonne National Laboratory, Lemont, Illinois 60439*
[2]*Department of Physics and Astronomy, Hofstra University, Hempstead, New York 11549*

We present a study of the superconducting properties of the candidate topological superconductor Ti$_3$Sb. Electrical transport measurements show zero resistance with a $T_{c,onset}$ of $\approx$ 5.9 K with a transition width $\Delta T_c \approx$ 0.6 K. The superconducting phase boundaries as derived from magneto-transport and magnetic susceptibility measurements agree well. We estimate an upper critical field $B_{c2}(0) \approx$ 4.5 T. A Ginzburg-Landau (GL) analysis yields values of the coherence length and penetration depth of ξ=6.2 nm and λ=340 nm, respectively, and a GL parameter $\kappa \approx$ 55, indicating extreme type-II behavior. Furthermore, we observed a step height in the specific heat $\frac{\Delta C_e}{\gamma T_c} \approx$ 1.61, a value larger than the Bardeen-Cooper-Schrieffer (BCS) value of 1.43, suggesting modest coupling. Measurements of the temperature dependence of the London penetration depth via the tunnel-diode oscillator (TDO) technique down to $\approx$ 450 mK show a full superconducting gap, consistent with a conventional *s*-wave gap structure.

[*]*rchapai@anl.gov*

# I. INTRODUCTION

Consideration of time-reversal symmetry, parity symmetry, and crystallographic symmetries in crystalline solids in recent years has revolutionized our understanding of condensed matter physics, leading to prediction and discovery of topological insulators [1], Dirac and Weyl semimetals [2, 3], and suggestions of Majorana modes in potential topological superconductors, among other discoveries [4, 5]. Even particles forbidden in high-energy physics such as triple and sextuple point fermions are believed to exist as quasiparticle excitations in topologically protected systems, with eightfold fermionic excitations predicted in several A15-structure intermetallic compounds [6-9]. These materials with topologically nontrivial electronic band structures display topological surface states with spin-polarized textures [10, 11]. A topological superconducting state may be induced into these surface states via proximity effect when the bulk of the material undergoes a transition into a conventional *s*-wave superconducting state. Thus, the presence of nontrivial topology in intrinsic superconducting materials offers the possibility of realizing topological superconductivity, obviating the complexity of fabricating a proximity-coupled heterostructure of a topological insulator and an *s*-wave superconductor [12].

The A15 class of materials has been studied for decades due to the importance of members such as $Nb_3Ge$ and $Nb_3Sn$ for the fabrication of high-field superconducting magnets [13-16]. However, their topological properties went unnoticed [10, 11] until recently, when it has now been proposed that A15 superconductors such as $Ta_3X$ (X= Sb, Sn, Pd) and $Nb_3Y$ (Y= Bi, Sb) have non-trivial topology [10, 11, 17, 18] and consequently should exhibit a large intrinsic spin Hall effect [11]. Non-trivial topology in superconducting A15 material thus provides an appealing platform to potentially realize topological superconductivity. As many of these compounds only received cursory examination in the 1960s and 1970s (and very little since), it is important to re-examine their bulk superconducting properties, particularly their gap structure.

Here we characterize the superconducting properties of one such A15 material, $Ti_3Sb$, through measurements of magneto-transport, magnetization, specific heat, and the temperature dependence of the London penetration depth, $\Delta\lambda(T)$. We find the onset of superconductivity at $\approx$ 5.9 K and lower and upper critical fields of 6.4 mT and 5 T, respectively. The gap ratio derived from specific heat anomaly, $2\Delta_0/k_B T_c = 3.53$, the value predicted by BCS theory when neglecting strong coupling effects. Adopting an electron-phonon coupling constant $\lambda_{ep} \sim 0.8$ obtained by density functional theory (DFT) [19], yields a gap ratio of 4.73, suggesting a modest coupling. Ginzburg-Landau (GL) analysis yields values of the coherence length and penetration depth of $\xi = 6.2$ nm and $\lambda = 340$ nm, respectively, and a GL parameter of $\kappa \sim 55$, indicating extreme type-II behavior. The observed temperature dependence of $\Delta\lambda$ indicates a full superconducting gap and is inconsistent with either point or line nodes in the gap.

## II. EXPERIMENTAL DETAILS

Polycrystalline Ti$_3$Sb was prepared from stoichiometric mixtures of titanium slugs (99.99% Alfa Aesar) and antimony shot (99.95% Alfa Aesar) by arc-melting in an argon atmosphere. To improve chemical homogeneity, the arc casted ingots were annealed at 900 ºC for one week in an evacuated fused silica tube. Phase purity and crystal structure of the alloy was determined with X-ray diffraction (XRD) measurements performed on a PANalytical X'pert Pro X-ray diffractometer with Cu K$_\alpha$ radiation, and the stoichiometry was verified via energy dispersive X-ray spectroscopy (EDX) measurement. The electrical resistivity was measured via a standard four-probe method. Thin gold wires (50 $\mu$m) were attached to a rectangular sample of ~2 mm length using Epotek H20E conductive silver epoxy. For magneto-transport measurements, rectangular shaped ~2 mm long samples broken off from the as-grown Ti$_3$Sb ingot were mounted on a custom-built probe in a Janis cryostat capable of 1.8 K and 9 T using a DC technique with a 1 mA current. Current direction was modulated at 1 Hz in a square wave form to mitigate thermal effects at wire junctions.

Magnetization measurements as a function of field and temperature were performed on a small irregular piece broken off from the as-grown ingot in a Quantum Design MPMS SQUID magnetometer at temperatures down to 1.8 K. Zero-field cooled (ZFC) warming and field cooled (FC) cooling measurements were performed. Heat capacity was measured in a Quantum Design Physical Properties Measurement System (PPMS) using relaxation method [20]. Tunnel diode oscillator (TDO) measurements [21] were performed down to ~450 mK in an Oxford $^3$He cryostat with a custom-built resonator [22, 23] running at approximately 14.5 MHz. The specimen was a small fragment (approximately 0.5 mm × 0.5 mm × 0.1 mm) cut from a larger piece pre-screened for superconductivity in the SQUID magnetometer. The TDO measures the temperature dependence of the London penetration depth $\Delta\lambda(T) = \lambda(T) - \lambda_0$, where $\lambda_0$ is the zero-temperature value.

## III. RESULTS AND DISCUSSION

Figure 1(a) shows the room temperature powder XRD pattern of $Ti_3Sb$ indexed in the A15 structure type ($Pm\overline{3}n$, space group 223) with a refined lattice parameter $a$=5.223(6) Å, which is consistent with previous reports [18, 24]. A small fraction of impurity phase is evident in the XRD data in Fig. 1(a), where the impurity peaks belonging to the tetragonal non-superconducting phase $Ti_5Sb_2$ [25] are marked with asterisks. EDX measurements (see Supplementary Material [26]) show an average atomic ratio of Ti:Sb $\approx$ 3.04:0.96, indicating slight Sb-deficiency. Superconducting $Ti_{1-x}Sb_x$ forms the A15 structure for x between 0.1 and 0.3 [27, 28], with the maximum $T_c$ found for x ~ 0.25, corresponding to $Ti_3Sb$. At lower Sb-content, $T_c$ decreases rapidly with decreasing x [28]. The temperature dependence of the electrical resistivity of $Ti_3Sb$ is shown in Fig. 1(b) from 390 K to 1.8 K, through the superconducting transition, with a $T_{c,onset}$ of ~5.9 K. The temperature dependence of the resistivity is in good agreement with that reported previously for $Ti_3Sb$ and is similar to other A15 materials [24, 27].

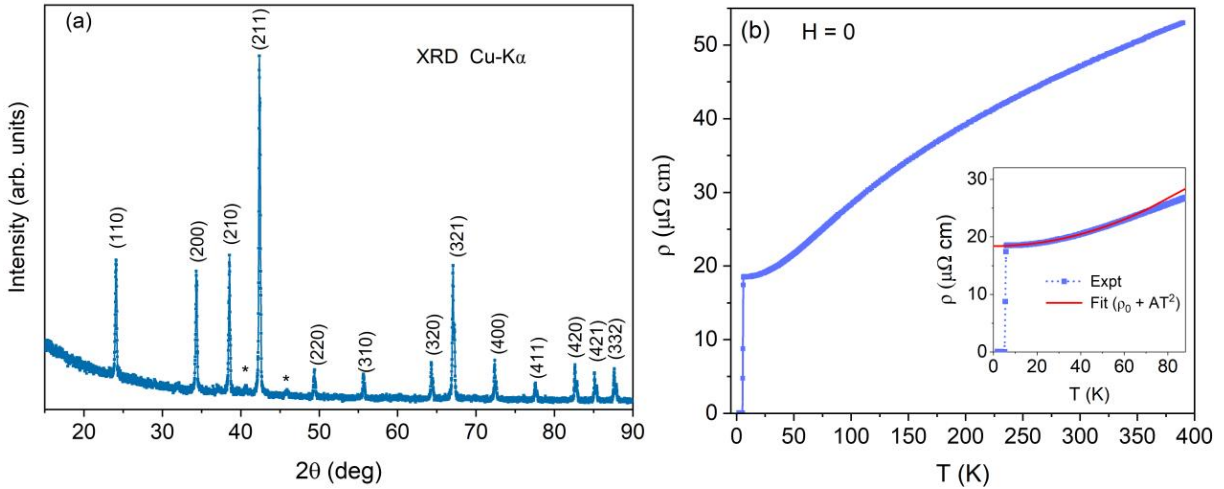

Figure 1. (a) Room temperature XRD pattern of $Ti_3Sb$ indexed in the A15 structure type ($Pm\overline{3}n$). The asterisks mark the peaks corresponding to traces of an impurity phase ($Ti_5Sb_2$). (b)Temperature dependence of electrical resistivity $\rho(T)$ of $Ti_3Sb$ in zero applied field; the onset temperature of superconductivity is ~5.9 K, and zero resistance is reached at ~5.35 K. The inset shows the $T^2$ dependence of the resistivity between $T_c$ and 60 K with a fit (red line).

The sample has an estimated residual resistivity ratio (RRR) of $\frac{\rho(300\ K)}{\rho(6\ K)} \approx 2.52$. Three samples were screened for resistivity, all samples measured had RRR values between 2.5 and 3. At temperatures below ~60 K (inset of Fig. 1(b)), the resistivity follows a $T^2$ dependence $\rho(T) =$ $\rho_0 + AT^2$ with $\rho_0 = 18.4\ \mu\Omega\ cm$ and $A = 0.00125\ \mu\Omega\ cm/K^2$, as has been reported before [24]. Such $T^2$-variation has been observed for many A15 compounds and various mechanisms have been proposed including non-Debye phonon spectra, non-conservation of momentum in electron-phonon scattering, and spin fluctuations [29-33]. In addition, in the Fermi liquid description of metals, electron-electron scattering yields a $T^2$-dependence at low temperatures. In fact, the Kadowaki-Woods ratio $\frac{A}{\gamma^2} = 0.7 \times 10^{-6} \mu\Omega\ cm\ \frac{mol\ f.u.^2 K^2}{mJ^2}$, where $\gamma = 42.19 \frac{mJ}{mol\ f.u.\ K^2}$ is the Sommerfeld coefficient of the electronic specific heat as determined below, falls well within the range typically seen for highly correlated transition metals and/or heavy Fermion materials [34-36].

Figure 2(a) shows the superconducting transition in more detail, as well as its continuous shift to lower temperature in magnetic fields applied in steps of 0.1 T from 0 T to 5 T. In zero field, the transition is sharp with small rounding at the top. As field increases, the transition slightly broadens. In 5 T, the transition is completely suppressed to below 1.8 K. Fig. 2(b) shows the transition as a function of applied magnetic field at various fixed temperatures between 1.8 K and 5.3 K. Figs. 2(a) and (b) reveal some broadening and the appearance of structure in the transition, particularly at high fields/low temperatures which may reflect a residual inhomogeneity in Sb-distribution and a corresponding variation in $T_c$ [27, 28]. No normal state magnetoresistance is observed in either set of measurements, consistent with the relatively high resistivity of the material. Using the midpoint of the resistive transition, both $\rho(T)$ and $\rho(H)$ measurement protocols yield essentially the same phase boundary of $H_{c2}(T)$, as discussed further below.

The zero-temperature London penetration depth, $\lambda_0$, and the coherence length, $\xi_0$, can be estimated from measurements of the lower and upper superconducting critical fields $H_{c1}$ and $H_{c2}$. $H_{c1}$ values were deduced from a series of low-temperature magnetization vs field measurements for a thin sample aligned with its length parallel to the field. The field dependence of magnetization was measured at fixed temperatures below $T_c$ following zero field cooling; the results are shown in Fig. 3(a).

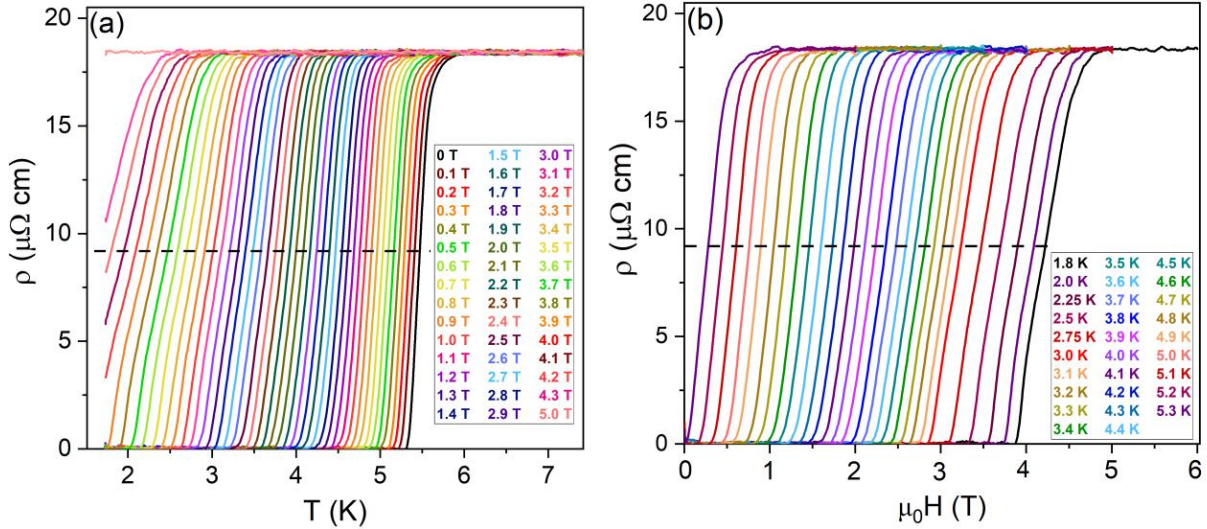

Figure 2. (a) Electrical resistivity of Ti$_3$Sb as a function of temperature in various applied magnetic fields. (b) The electrical resistivity as a function of magnetic field at various fixed temperatures from 1.8 K to 5.3 K. The dashed lines mark the definition of $T_c$ as the midpoint of the transition.

By fitting the $M$ versus $H$ data with a straight line in the low-field region and locating the field at which $M(H)$ deviates from the linear Meissner effect, we obtain the penetration field, $H_p$, at which vortices first enter the sample, see inset in Fig. 3(b). This is related to the lower critical field $H_{c1}$ via the demagnetization coefficient $N$. Since the sample surface is rough, we do not consider effects due to surface barriers here [37-40]. By comparing the low-field slope of $M(H)$ at the lowest temperature to that of an ideal diamagnet, $-\frac{1}{4\pi}$, we observe that the demagnetization factor of our sample is small ($N = 0.035$), leading to only minor corrections to the values of $H_{c1}$: $H_{c1} = \frac{1}{1-N}H_P$ as presented in Fig. 3(b). The red line in the figure represents a phenomenological parabolic temperature dependence $H_{c1} = H_{c1}(0)(1 - t^2)$ with $t = T/T_c$ to the $H_{c1}$ datapoints, from which we extrapolate $H_{c1}$ to be approximately 6.4 mT, a value significantly smaller than reported previously [18]. We attribute this discrepancy in $H_{c1}$ to a distribution of $T_c$ brought about by the inhomogeneous Sb-concentration. As evidenced by the inset of Fig. 3(b), our measurements probe the lowest value of $H_{c1}$ in the sample, while resistivity measurements probe the highest value of $T_c$ (at which a superconducting filament can be established across the sample). Correspondingly, the data in Fig. 3(b) seem to extrapolate to lower value of $T_c$ than seen in the transport measurements.

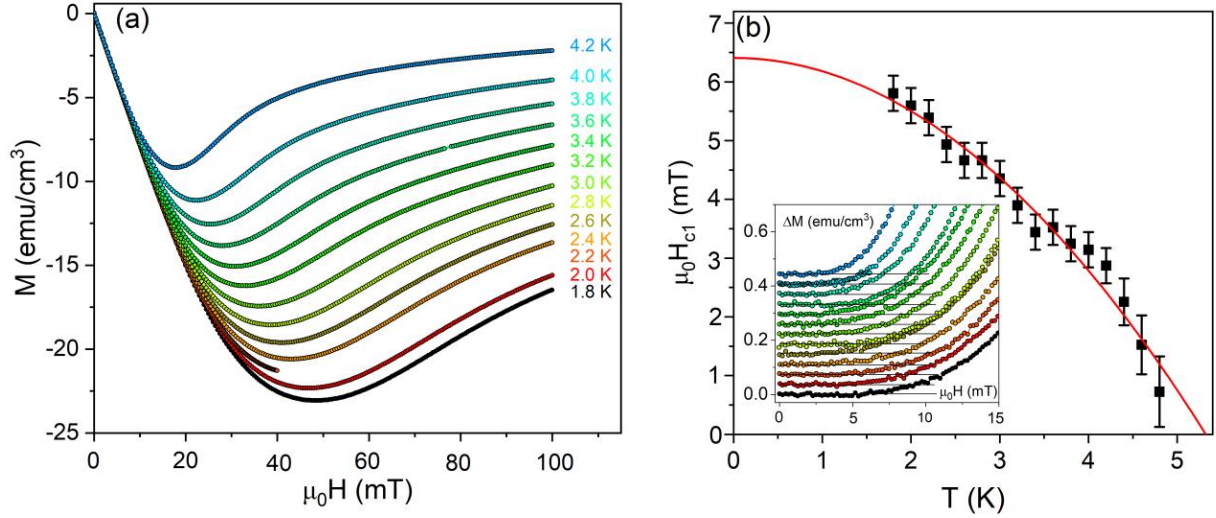

Figure 3. (a) Zero field-cooled magnetization vs applied field at multiple temperatures. The penetration field, taken as approximately equal to $H_{c1}$, is taken as the field where $M(H)$ first deviates from linearity. (b) $H_{c1}$ values obtained from the data in frame (a). The red line is a phenomenological fit (see text) with an extrapolation to the zero-temperature value. Inset: $\Delta M$ vs $H$ indicating criterion for the determination of $H_{c1}$.

Figure 4(a) shows the temperature dependence of the magnetization measured upon cooling in multiple fields. A Curie-Weiss fit above $T_c$ was subtracted from each dataset, and the onset of superconductivity, marked by vertical lines, is shown in the upper inset for each field. A ZFC magnetization measurement, obtained upon warming in an applied field of 10 mT, is shown in the lower inset which yields $T_{c,onset} \sim 5.5$ K while the extrapolation of the steepest slope yields ~4.8 K reflecting the distribution of $T_c$-values. The irreversibility line $H_{irr}(T)$ of a superconductor separates reversible and irreversible regions in the superconducting phase diagram; non-zero critical currents are only found below the irreversibility line as it marks the onset of effective vortex pinning. The irreversibility line may be determined from magnetization hysteresis measurements as the field at fixed temperature at which there is an onset of superconducting hysteresis, or via a comparison of ZFC and FC magnetization measurements at the temperature in fixed field at which the ZFC and FC measurements diverge. The ZFC and FC magnetization data for Ti$_3$Sb is shown in Fig. 4(b) in an applied field of 1 T, with the inset showing the splitting of the ZFC and FC curves at this field. The irreversibility temperature is marked with a blue dashed line. Multiple ZFC and FC traces in different fields were acquired and analyzed to determine the irreversibility line, shown further below with the superconducting phase boundary line. Magnetic hysteresis measurements performed at fixed temperatures are shown in Fig. 4(c), from which it is possible to extract the irreversibility field. The hysteresis loops are symmetric indicating that bulk pinning, rather than surface pinning, is the dominant pinning mechanism. Furthermore, the results show that the magnetization hysteresis, that is, the superconducting critical current density of Ti$_3$Sb, is rapidly suppressed in fields of the order of 1 T.

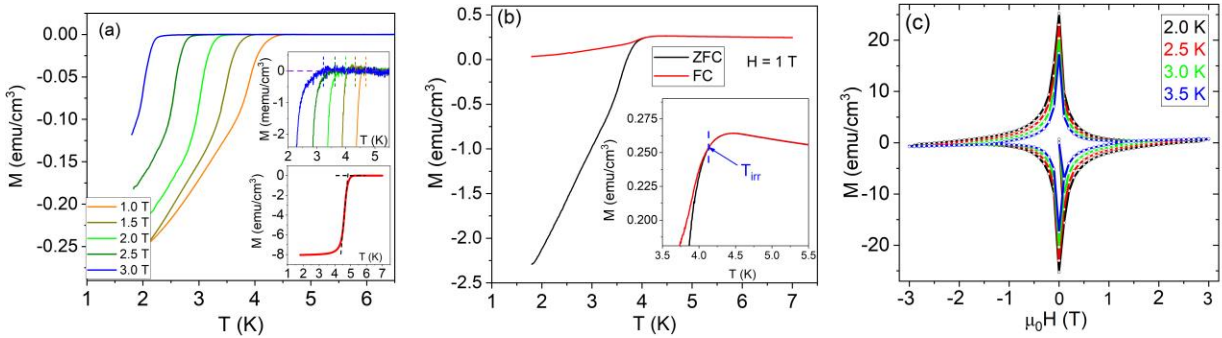

Figure 4. (a) FC magnetization vs temperature following background subtraction, measured on cooling. Upper inset: the onset of superconductivity in each value of applied field. Lower inset: ZFC magnetization vs temperature in an applied field of 10 mT, measured upon warming. (b) ZFC (black) and FC (red) magnetization vs temperature in an applied field of 1 T, measured upon warming. The inset shows a zoom-in where the ZFC and FC curves separate (blue dashed line) at the irreversibility temperature $T_{irr}$ at this field. (c) Magnetic hysteresis measurements at various temperatures below $T_c$.

The sample was further characterized by calorimetric measurements. Figure 5(a) presents the temperature dependence of the specific heat, $C(T)$, of Ti$_3$Sb between 2 and 200 K. $C(T)$ approaches the classical value of $3NR$ at high temperature, decreases with decreasing temperature and displays a clear anomaly at $T_c$ ~5 K (see inset: Fig. 5(a)) signaling the onset of bulk superconductivity. The anomaly in $C(T)$ can be completely suppressed with the application of a field of 6 T (see inset: Fig 5(a)) consistent with transport measurements (Fig. 2). The onset of the transition in calorimetry is slightly lower than that observed in transport or magnetization, consistent with calorimetry probing the entire bulk of the sample, whereas transport is sensitive to filamentary paths. To quantitatively analyze $C(T)$, we plot data as $C/T$ versus $T^2$ at low temperatures as shown Fig. 5(b). Fitting the normal state data above $T_c$ with $\frac{C(T)}{T} = \gamma + \beta T^2$ (represented by the solid line in Fig. 5(b)) yields the Sommerfeld coefficient $\gamma = 42.19(5)$ mJ mol$^{-1}$K$^{-2}$, a value close to that reported earlier [18] and $\beta = 0.975(1)$ mJ mol$^{-1}$K$^{-4}$. Using the Debye relation $\beta = N\left(\frac{12}{5}\right)R\pi^4\Theta_D^{-3}$, we obtain a Debye temperature $\Theta_D \sim 200$ K for Ti$_3$Sb. This value of $\Theta_D$ is similar to that reported for other A15s [41-43].

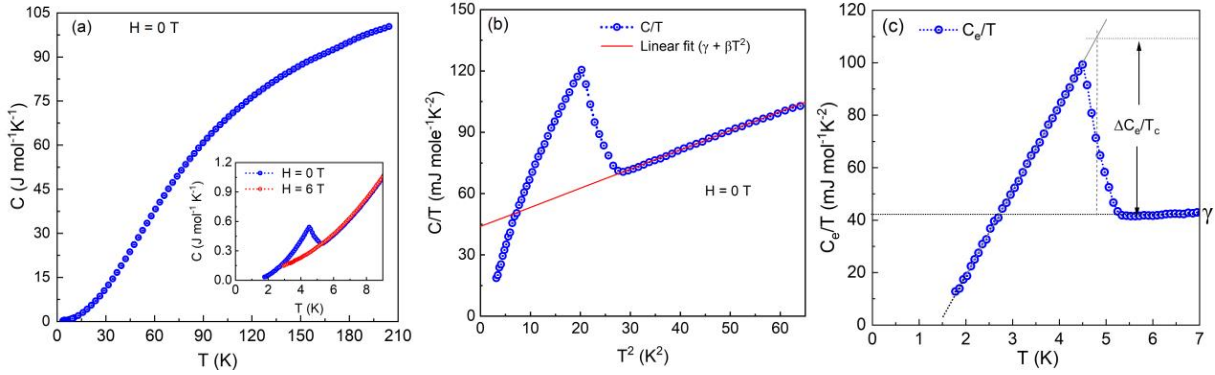

Figure 5. (a) Temperature dependence of the specific heat $C(T)$ of Ti$_3$Sb between 2 and 200 K. Inset: $C(T)$ between 2 and 9 K at indicated applied fields. (b) $C/T$ versus $T^2$. The solid line is a linear fit of the data to the relation $\frac{C}{T} = \gamma + \beta T^2$. (c) Electronic specific heat plotted as $C_e/T$ versus $T$. The $T_c$ is determined by approximating the measured data with an ideal step and an entropy conserving construction, as indicated by dotted lines in the figure.

By subtracting the lattice contribution ($\beta T^2$) from the total specific heat, the electronic part of specific heat divided by temperature, $C_e/T$, is plotted as a function of $T$ in Fig. 5(c). We approximate the broadened transition, indicating $T_c$ values ranging from ~ 4.5 K to ~ 5.5 K, by an ideal step using an entropy-conserving construction as indicated by the dashed line in Fig. 5(c) yielding $T_c$ = 4.8 K, consistent with the magnetization data in Fig. 4. The corresponding values of the step in the specific heat are $\Delta C_e/T_c$ = 68.5 mJ mol$^{-1}$K$^{-2}$ and $\Delta C_e/\gamma T_c$ ~ 1.61 which is slightly larger than the BCS value of 1.43 expected for a weak-coupling, single-band, isotropic *s*-wave superconductor [44]. A similar result is obtained from the normalized jump of the slope of the specific heat, $(T_c/\Delta C)\Delta(dC/dT)|_{T_c}$. In single-band weak coupling BCS theory, this ratio is 2.64, while in the case of strong-coupling such as in Pb it is 4.6 [44] and in the two-band superconductor MgB$_2$, a value of 3.35 can be deduced [45]. From the data in Fig. 5 we obtain a value of $(T_c/\Delta C)\Delta(dC/dT)|_{T_c} \sim 2.75$, slightly above the weak-coupling limit.

At present, our specific heat data are limited to 1.8 K as the lowest temperature, precluding determination of the energy gap structure from fits of the temperature dependence of the specific heat. Nevertheless, we may estimate the condensation energy, $U(0)$ from the integral of the specific heat as $U(0) = \int_0^{T_c}(C_s - C_n)dT$ yielding $U(0)$ ~ 232.34 mJ/mol-f.u, where $C_s$ and $C_n$ are specific heat in the superconducting and normal state respectively. Here, we assume that residual contributions to the electronic specific heat, for instance due to the Ti$_5$Sb$_2$ phase, are negligible, and that between 0 K and 1.8 K the temperature dependence of $(C_s - C_n)$ is linear. Errors due to the latter will have negligible effect on the total integral. From the condensation energy, the thermodynamic critical magnetic field $B_c(0)$ can be estimated via $U(0) = -B_{c(0)}^2/2\mu_0$ (where $\mu_0$ is the vacuum permeability) yielding in $B_c(0)$ = -0.117 T. Furthermore, from the Rutgers relation $\frac{\Delta C}{T_c} = \frac{V_m}{\mu_0}\left(\frac{\partial B_c}{\partial T}\right)^2_{T_c}$ ($v_m \approx 42.8$ cm$^3$ is the molar volume) $-\partial B_c/\partial T|_{T_c} = 44.8$ mT/K is obtained. We then estimate the reduced thermodynamic critical field $-\frac{B_c(0)}{\left(\frac{\partial B_c}{\partial T}\right)_{T_c} T_c} = 0.544$, a value slightly lower than the BCS value of 0.576 for a weak-coupling single-band isotropic $s$-wave superconductor [44]. We note that strong-coupling effects reduce the value of the reduced thermodynamic critical field. The condensation energy can be expressed in terms of microscopic parameters $N(0)$ and $\Delta_0$ through $U(0) = \frac{N(0)\Delta_0^2}{2}$, where $N(0)$ is the electronic density of states per spin at the Fermi level and $\Delta_0$ is the zero-temperature superconducting gap. The density of states (per spin) is related to the Sommerfeld coefficient of the electronic specific heat as $\gamma = 2\pi^2 k_B^2 v_m N(0)\frac{(1+\lambda_{ep})}{3}$, where $\lambda_{ep}$ is the electron-phonon coupling constant. While our above analysis of the specific heat data is consistent with weak or modest coupling strength, i.e., $\lambda_{ep} \ll 1$, a recent DFT calculation [19] predict strong coupling, $\lambda_{ep} \sim 0.8$, for Ti$_3$Sb with $T_c \sim 6.4$ K. We note that neglecting strong coupling effects yields a gap value of $\Delta_0 = 0.73$ meV and a gap ratio of $2\Delta_0/k_B T_c = 3.53$, the value expected from BCS theory and consistent with the above discussions. Adopting $\lambda_{ep} = 0.8$ would increase the gap ratio to 4.73.

We employ the TDO technique to trace the superconducting transition and to map the temperature dependence of the penetration depth down to 0.45 K, corresponding to $T/T_c \sim 0.1$. The shift of the TDO frequency with temperature and/or field is a measurement of the degree of screening of magnetic flux in the sample, which above $T_c$ is due to the normal-state skin depth and below $T_c$ is due to superconducting flux screening [21]. In a TDO measurement, the transition into or out of the superconducting state is accompanied by a large shift in oscillator frequency proportional to the magnetic susceptibility, allowing mapping of the temperature dependence of the upper critical field of the superconducting state. TDO measurements on a roughly plate-like piece are shown in Fig. 6(a) in applied magnetic field values of up to 1 T in increasing step sizes. The onset of the transition in zero field is approximately 5.85 K, consistent with that observed in transport and magnetization measurements. With increasing field, the diamagnetic transition onset is shifted to lower temperatures. There is a foot visible in the TDO data, extending between 4.5 and 5 K, likely caused by inhomogeneity of the Sb-concentration as discussed above. This feature is quickly subsumed by the bulk transition in increasing field. In all measurements, the sample was field cooled from above $T_c$, then data were collected during a slow warming ramp through the transition and beyond. As the resonant frequency of the TDO circuit itself is sensitive to magnetic fields, the presented curves were all offset in frequency by constant amounts such that the TDO responses in the normal state aligned. To establish a consistent $T_c$ criterion between measurement techniques, we use the following procedure. First, we find the $\Delta f$ decrease from the normal state, temperature independent behavior in zero applied field at the temperature equal to $T_{c,midpoint}$ from transport measurements (5.74 K), then declare this $\Delta f$ our criterion for $T_c$ in all applied fields in TDO measurements. This approach forces $T_c$ from both TDO and transport in zero field to be identical and gives remarkably consistent agreement between superconducting phase boundaries from each technique as shown below in Fig. 6(b).

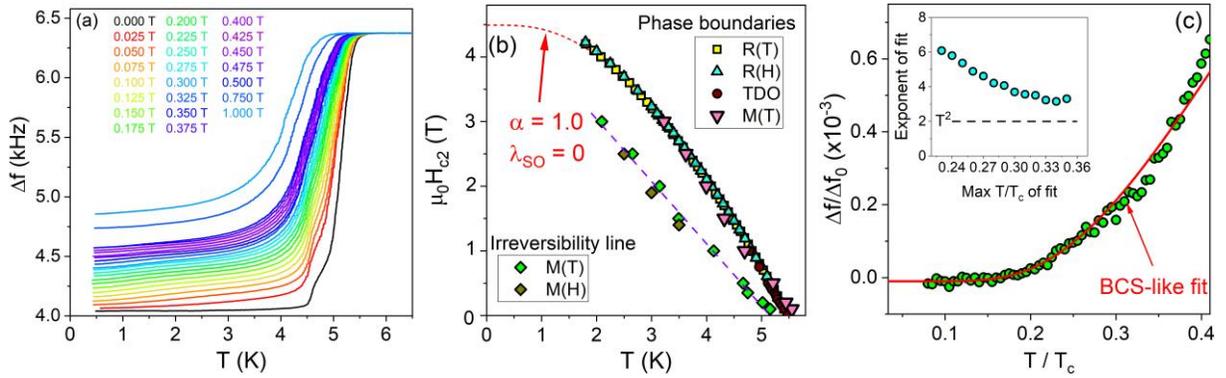

Figure 6. (a) The TDO frequency shift vs temperature of Ti$_3$Sb in multiple applied field values. (b) Superconducting phase diagram for Ti$_3$Sb from magneto-transport (yellow, cyan), TDO (maroon), and magnetization (pink) measurements, including an estimate of the irreversibility line (purple) from magnetization (green, dark yellow) measurements. The red dashed line is a WHH model fit to the phase boundary data with no spin-orbit coupling. (c) Normalized low-temperature frequency shift $\Delta f(T)$. The inset shows the exponent of a $T^n$ fit to the data vs reduced temperature cutoff for the fit.

The magneto-transport, magnetization, and TDO measurements can be used to determine a superconducting phase diagram for Ti$_3$Sb, including the irreversibility line in the superconducting state. The results are shown in Fig. 6(b); multiple transport samples had virtually identical phase boundaries. The phase boundaries as derived from resistivity measurements in fixed fields as a function of temperature (yellow squares), at fixed temperatures as a function of field (cyan triangles), via TDO measurements (maroon circles), and via SQUID magnetometry measurements (pink triangles); all agree well with each other. Via the relation $B_{c2}^{GL} = \Phi_0/(2\pi\xi_{GL}^2)$, where $\Phi_0$ is the flux quantum, and $B_{c2}^{GL} = -(dB_{c2}/dT)|_{T_c}T_c = 8.5$ T is the zero-temperature GL upper critical field, we calculate a zero-temperature GL coherence length $\xi_{GL}$ of approximately 6.2 nm. The extrapolated zero-temperature value of the experimental $B_{c2}$ is 4.5 T comparable to the results in [18]. This extrapolated value amounts to about half the paramagnetic limiting field, which the BCS theory, has a value of $\mu_0 H_{c2}^{Pauli} = 1.86 \times T_c \sim 10\ T$, indicating that paramagnetic limiting, in addition to orbital limiting, contributes to the measured $B_{c2}$. This finding is consistent with the observation that $B_{c2}$ is lower than the orbitally limited upper critical field as given in WHH theory, $B_{c2}^{orbital} = 0.69 \times \mu_0 H_{c2}^{GL} \sim 5.86\ T$ [46]. As Ti$_3$Sb contains only light atoms, spin-orbit scattering is not expected to be effective. Then, the actual upper critical field in the dirty limit is given as $B_{c2} = B_{c2}^{orbital}/\sqrt{1+\alpha_M^2}$, where the Maki parameter $\alpha_M = \sqrt{2}B_{c2}^{orbital}/B_{c2}^{Pauli}$ measures the relative role of orbital and paramagnetic limiting [47]. These results yield relatively large values for the Maki parameter $\alpha_M \sim 0.8$ while a fit of the data to the WHH expression [26, 46] returns $\alpha_M = 1$ and $B_{c2}(0) = 4.5$ T (see Fig. 6(b)). We note though that the expression for the paramagnetic limit as well as the WHH expression are based on the BCS weak-coupling formalism [48] which would lead to an overestimation of $\alpha_M$. For instance, $B_{c2}^{Pauli}$ contains an additional factor of $(1+\lambda_{ep})^n$ where $n$ is 0.5 [49] or 1 [50, 51].

Additionally, using the GL relation $H_{c1} = (\Phi_0/(4\pi\lambda^2))(\ln[\lambda/\xi] + 0.5)$, and the GL coherence length deduced above we estimate the zero-temperature London penetration depth $\lambda_0$ to be approximately 340 nm. This value implies a GL parameter $\kappa \sim 55$, confirming Ti$_3$Sb to be extreme type-II. Alternatively, $\kappa$ can be obtained from the calorimetric data via $dB_{c2}/dT|_{T_c} = \sqrt{2}\ \kappa\ dB_c/dT|_{T_c}$ yielding $\kappa \sim 24$ and $\lambda_0 \sim 150$ nm. We attribute the difference in these estimates to the presence of a distribution of Sb-concentration and the corresponding variation in $T_c$. As mentioned above, this variation affects measured quantities in different ways. While magnetization measurements (Fig. 3) highlight the lowest value of $B_{c1}$, resistivity measurements accentuate the highest $B_{c2}$ and specific heat measurements yield a volume average (Fig. 5). Thus, our results suggest that with Sb-deficiency the penetration depth increases strongly.

Low-temperature penetration depth measurements were carried out via the TDO technique in the temperature range of 0.45 K to 10 K to investigate the superconducting gap structure. In the TDO technique, the frequency shift $\Delta f$ of the resonator is proportional to the change of the penetration depth [21, 52]:

$$\Delta f(T) = G\Delta\lambda(T)$$

where the geometrical factor G depends on the geometry of the resonator coil as well as the volume and shape of the sample. As the ac field in the coil has a very small amplitude (~2 $\mu T$), the sample remains fully in the Meissner state below $T_c$. The low-temperature variation of the London penetration depth can provide information on the superconducting gap structure [21]. At sufficiently low temperature $\left(\frac{T}{T_c} < 0.33\right)$, conventional BCS theory for an isotropic *s*-wave superconductor yields an exponential temperature dependence in $\Delta\lambda(T)$ [21]:

$$\frac{\Delta\lambda(T)}{\lambda_0} = \sqrt{\frac{\pi\Delta_0}{2T}}\exp\left(\frac{-\Delta_0}{T}\right)$$

where $\Delta_0$ and $\lambda_0$ are the zero-temperature values of the superconducting energy gap and the penetration depth. In a nodal superconductor a stronger temperature dependence is observed due to enhanced thermal excitation of quasiparticles near the gap nodes, where the gap amplitude is suppressed. As a result, there is a power-law variation in the penetration depth, $\Delta\lambda \sim T^n$, where the exponent $n$ depends on the nature of the nodes and the degree of electron scattering. Line nodes in the energy gap will generate a $T$-linear response, while point nodes will generate a $T^2$-response [53]. Behavior with $n \geq 3$ is generally considered indistinguishable from exponential and is taken as evidence of a full superconducting gap.

The evolution of the normalized low-temperature TDO frequency shift of Ti$_3$Sb is shown in Fig. 6(c). Instrument noise is on the order of 0.25 Hz, and with averaging slow temperature sweeps back and forth between 0.24 K and 2 K, a noise level of < 0.05 Hz is achievable, with a full superconducting transition frequency shift of approximately 2300 Hz. Below $\frac{T_c}{3}$, the variation of the frequency shift is best described (red line in Fig. 6(c)) by an exponential dependence with a BCS-like gap value. The observed gap ratio of $\frac{\Delta_0}{k_B T_C} = 1.23$ is smaller than the standard BCS value $\frac{\Delta_0}{k_B T_C} = 1.76$ for an isotropic single-gap *s*-wave superconductor [44]. However, similar behavior has been reported for Nb$_{1-x}$Sn$_x$ [54] where $T_c$ and the gap decrease rapidly as $x$ decreases below 0.25 such that for $x < 0.24$ the apparent gap ratio falls below the week coupling limit, a behavior attributed to effects due to inhomogeneous Sn-content [54]. The inset of Fig. 6(c) shows the exponent in a fit of the form $AT^n$ with the exponent $n$ a free parameter, while the fit is calculated from the minimum temperature to various cutoff values of reduced temperature $t = T/T_c$. Over the entire low-temperature range in the data relevant to gap structure analysis, the fitted exponent $n$ is at least 3, and as the maximum window decreases the fit exponent grows, excluding line nodes (which would present as $T$-linear) and point nodes (which would present as $T^2$), leaving the conclusion that the gap, while possibly anisotropic or non-uniform across the sample, is certainly nodeless. This result is not affected by the presence of a distribution of gaps since the measurement temperature is low enough such that $k_B T \ll \Delta_{min}$, and the measured temperature dependence is exponential.

## IV. SUMMARY

In summary, we have systematically characterized the superconducting properties of polycrystalline Ti$_3$Sb, an A15 compounds, using magneto-transport, specific heat, DC, and AC susceptibility measurements. We estimate a zero-temperature upper critical field of ~4.5 T and a slope at $T_c$ of -1.54 T/K, corresponding to a GL coherence length of 6.2 nm, and find that the material is extreme type-II. Considerable curvature away from conventional orbital limiting behavior in the superconducting phase diagram indicates non-negligible paramagnetic limiting and α$_M$ ~ 1. Furthermore, the step height in the specific heat $\Delta C_e/\gamma T_c$ ~ 1.61 and the reduced thermodynamic critical field of 0.54 imply modestly strong coupling. An exponential temperature dependence is observed in the London penetration depth Δλ indicating a nodeless superconducting gap. This result holds true even in the presence of inhomogeneous Sb-content, implying that all superconducting variants contained in the sample are fully gapped with gaps much larger than 0.4 K, the lowest temperature in the penetration depth measurements. Our current results on polycrystalline samples do not give information on whether the predicted topological band structure contributes to superconductivity and Hall effect in Ti$_3$Sb. We envision that Ti$_3$Sb is in the same group of materials as FeSe$_x$Te$_{1-x}$ in which *s*-wave superconductivity in the bulk induces topological surface superconductivity [10] that can be revealed in spectroscopic techniques such as scanning tunneling microscopy (STM) and angle resolved photoemission spectroscopy (ARPES) once single crystal sample become available.

## Acknowledgments

This work was supported by the U. S. Department of Energy, Office of Science, Basic Energy Sciences, Materials Sciences and Engineering Division. MPS acknowledges support from the U.S. Department of Energy, Office of Science, Office of Workforce Development for Teachers and Scientists (WDTS) under the Visiting Faculty Program (VFP).

## References

1. L. Fu, C. L. Kane, and E. J. Mele, Topological insulators in three dimensions, *Phys. Rev. Lett.* **98**, 106803 (2007).
2. Z. K. Liu, B. Zhou, Y. Zhang, Z. J. Wang, H. M. Weng, D. Prabhakaran, S.-K. Mo, Z. X. Shen, Z. Fang, X. Dai *et al*., Discovery of a Three-Dimensional Topological Dirac Semimetal, Na$_3$Bi, *Science* **343,** 864 (2014).
3. B. Q. Lv, H. M. Weng, B. B. Fu, X. P. Wang, H. Miao, J. Ma, P. Richard, X. C. Huang, L. X. Zhao, G. F. Chen *et al*., Experimental Discovery of Weyl Semimetal TaAs, *Phys. Rev. X* **5**, 031013 (2015).
4. V. Mourik, K. Zuo, S. M. Frolov, S. R. Plissard, E. P. A. M. Bakkers, and L. P. Kouwenhoven, Signatures of Majorana Fermions in Hybrid Superconductor-Semiconductor Nanowire Devices, *Science* **336**, 1003 (2012).
5. C. Beenakker and L. Kouwenhoven, A road to reality with topological superconductors, *Nat. Phys*. **12**, 618 (2016).
6. Q. Liu and A. Zunger, Predicted realization of cubic Dirac Fermion in quasi-one-dimensional transition-metal monochalcogenides, *Phys. Rev. X*. **7**, 021019 (2017).
7. B. Bradlyn, J. Cano, Z. Wang, M. G. Vergniory, C. Felser, R. J. Cava, and B. A. Bernevig, Beyond Dirac and Weyl fermions: Unconventional quasiparticles in conventional crystals, *Science* **353**, aaf5037 (2016).

8. R. Chapai, Y. Jia, W. A. Shelton, R. Nepal, M. Saghayezhian, J. F. DiTusa, E. W. Plummer, C. Jin, and R. Jin, Fermions and bosons in nonsymmorphic PdSb$_2$ with sixfold degeneracy, *Phys. Rev. B* **99**, 161110(R) (2019).

9. X. Yáng, T. A. Cochran, R. Chapai, D. Tristant, J-X. Yin *et. al.*, Observation of sixfold degenerate fermions in PdSb$_2$, *Phys. Rev. B* **101**, 201105(R) (2020).

10. M. Kim, C.-Z. Wang, and K.-M. Ho, Topological states in A15 superconductors, *Phys. Rev. B* **99**, 224506 (2019).

11. E. Derunova, Y. Sun, C. Felser, S. S. P. Parkin, B. Yan, and M. N. Ali, Giant intrinsic spin Hall effect in W$_3$Ta and other A15 superconductors, *Sci. Adv*. **5**, eaav8575 (2019).

12. B. Jäck, Y. Xie, J. Li, S. Jeon, B. A. Bernevig, A. Yazdani, Observation of a Majorana zero mode in a topologically protected edge channel, *Science* **364**, 1255 (2019).

13. A. Echarri and M. Spadoni, Superconducting Nb$_3$Sn: A review, *Cryogenics* **11**, 274 (1971).

14. J. Gavaler, M. Janocko, A. Braginski and G. Roland, Superconductivity in Nb$_3$Ge, *IEEE Transactions on Magnetics*, **11**, 192 (1975).

15. A. Godeke, A review of the properties of Nb$_3$Sn and their variation with A15 composition, morphology and strain state, *Supercond. Sci. Technol*. **19** R68 (2006).

16. A. Godeke, Performance Boundaries in Nb$_3$Sn Superconductors, Ph.D. thesis, University of Twente, Enschede, The Netherlands, ISBN 90-365-2224-2 (2005), and references therein.

17. Q. Chen, Y. Zhou, B. Xu, Z. Lou, H. Chen et al., Large Magnetoresistance and Nontrivial Berry Phase in Nb$_3$Sb Crystals with A15 Structure, *Chin. Phys. Lett.*, **38**, 087501 (2021).

18. M. Mandal, K. P. Sajilesh, R. R. Chowdhury, D. Singh, P.K. Biswas, A.D. Hillier, and R. P. Singh, Superconducting ground state of the topological superconducting candidates Ti$_3$X (X= Ir, Sb), *Phys.* Rev. B **103**, 054501 (2021).

19. C. Tayran, M. Kim and M. Çakmak, Electron–phonon coupling of the Ti$_3$Sb compound, *J. Appl. Phys*. **132**, 075103 (2022).

20. G. R. Stewart, Measurement of low-temperature specific heat, *Rev. Sci. Instrum*. **54**, 1 (1983).

21. R. Prozorov and R. W. Giannetta, Magnetic penetration depth in unconventional superconductors, *Supercond. Sci. Technol*. **19**, R41 (2006).

22. M. P. Smylie, H. Claus, U. Welp, W.-K. Kwok, Y. Qiu, Y. S. Hor, and A. Snezhko, Evidence of Nodes in the Order Parameter of the Superconducting Doped Topological Insulator $Nb_xBi_2Se_3$ via Penetration Depth Measurements, *Phys. Rev. B* **94**, 180510 (2016).

23. M. P. Smylie, M. Leroux, V. Mishra, L. Fang, K. M. Taddei, O. Chmaissem, H. Claus, A. Kayani, A. Snezhko, U. Welp, and W.-K. Kwok, Effect of proton irradiation on superconductivity in optimally doped $BaFe_2(As_{1-x}P_x)_2$ single crystals, *Phys. Rev. B* **93**, 115119 (2016).

24. S. Ramakrishnan and G. Chandra, Resistivity studies of low $T_c$ A15 compounds, *Phys. Rev. B* **38**, 9245 (1988).

25. H. Auer-Welsbach, H. Nowotny et A. Kohl, Monatsh. *Chem.* **89**, 155 (1958).

26. Supplemental information includes energy dispersive X-ray spectroscopy (EDX) spectrum and WHH analysis incorporating orbital limiting, Pauli paramagnetic limiting, and spin-orbit coupling.

27. S. Ramakrishnan and G. Chandra, Normal state resistivity and $T_c$ studies of superconducting $Ti_{1-x}Sb_x$ system, *Phys. Lett. A* **100**, 441 (1984).

28. A. Junod, F. Heiniger, J. Muller, and P. Spitzli, Superconducting and specific heat in $Ti_3Sb$-based alloys. *Helv. Phys. Acta* **43**, 59 (1970).

29. G. W. Webb, Z. Fisk, J. J Engelhardt and S. D. Bader, Apparent $T^2$ dependence of the normal-state resistance and lattice heat capacities of high $T_c$-superconductors, *Phys. Rev. B* **15**, 2624 (1977).

30. R. Hasegawa, Electrical resistivity of amorphous metallic alloys, *Phys. Lett.* A **36**, 425 (1971).

31. M. Gurvitch, Universal disorder-induced transition in the resistivity behavior of strongly coupled metals, *Phys. Rev. Lett*. **56**, 647 (1986).

32. S. Ramakrishnan and G. Chandra, Comments on "Universal disorder-induced transition in the resistivity behavior of strongly coupled metals", *Phys. Rev. Lett*. **57**, 1961 (1986).

33. M. Kaveh and N. Wiser, Electron-electron scattering in conducting metals, *Adv. Phys*. **33**, 257 (1984).

34. K. Miyake, T. Matsuura, C. M. Varma, Relation between resistivity and effective mass in heavy Fermion and A15 compounds, *Solid State Commun*. **71**, 1149 (1989).

35. K. Kadowaki, S. B. Woods, Universal Relationship of the resistivity and specific heat in heavy fermion compounds, *Solid State Commun*. **58**, 507 (1986).

36. A. C. Jacko, J. O. Fjarestadt, B. J. Powell, A unified explanation of the Kadowaki-Woods ratio in strongly correlated metals, *Nat. Phys*. **5**, 423 (2009).

37. B. Shen, M. Leroux, Y. L. Wang, X. Luo, V. K. Vlasko-Vlasov, A. E. Koshelev, Z. L. Xiao, U. Welp, W. K. Kwok, M. P. Smylie, A. Snezhko, and V. Metlushko, *Phys. Rev. B* **91**, 174512 (2015).

38. C. P. Bean and J. D. Livingston, Surface barrier in type-II superconductors, *Phys. Rev. Lett.* **12**, 14 (1964).

39. M. Benkraouda and J. R. Clem, Magnetic hysteresis from the geometrical barriers in type-II superconducting strips, *Phys. Rev. B* **53**, 5716 (1996).

40. E. H. Brandt, G. P. Mikitik, and E. Zeldov, Two regimes of vortex penetration into platelet-shaped type-II superconductors, *J. Exper. Theor. Phys*. **117**, 439 (2013).

41. G. R. Stewart, Superconductivity in the A15 structure, *Phys. C* **514**, 28 (2015).

42. A. Junod, T. Jarlborg, J. Muller, Heat-capacity analysis of a large number of A15-type compounds, *Phys. Rev. B* **27**, 1568 (1983).

43. R. Chapai, A. Rydh, M. P. Smylie, D. Y. Chung, H. Zheng, A. E. Koshelev, J. E. Pearson, W.-K. Kwok, J. F. Mitchell, U. Welp, Superconducting properties of a spin Hall candidate $Ta_3Sb$ with eightfold degeneracy, *Phys. Rev*. B **105**, 184510 (2022).

44. J. P. Carbotte, Properties of boson-exchange superconductors, *Rev. Mod. Phys*. **62**, 1027 (1990).

45. F. Bouquet, R. A. Fisher, N. E. Phillips, D. G. Hinks, J. D. Jorgensen, Specific heat of $Mg^{11}B_2$: Evidence for a second energy gap, *Phys. Rev. Lett*. **87**, 047001 (2001).

46. N. R. Werthamer, E. Helfand, and P. C. Hohenberg, Temperature and Purity Dependence of the Superconducting Critical Field, $H_{c2}$. III. Electron Spin and Spin-Orbit Effects, *Phys. Rev*. **147**, 295 (1966).

47. K. Maki, Effect of Pauli paramagnetism on magnetic properties of high-field superconductors, *Phys. Rev*. **148**, 362 (1966).

48. C.T. Rieck, K. Scharnberg and N. Schopohl, Quasiclassical theory of the upper critical field of high-field superconductors. Application to momentum-dependent scattering, J. *Low Temp. Phys*. **84**, 381 (1991).

49. M. Schossmann and J. P. Carbotte, Pauli limiting of the upper critical magnetic field, *Phys. Rev. B* **39**, 4210 (1989).
50. D. Rainer, G. Bergmann and U. Eckhardt, Strong-coupling calculation of the upper critical field of amorphous lead, *Phys. Rev. B* **8**, 5324 (1973).
51. T. P. Orlando, E. J. McNiff, Jr., S. Foner, and M. R. Beasley, Critical fields, Pauli paramagnetic limiting, and materials parameters of $Nb_3Sn$ and $V_3Si$, *Phys. Rev. B* **19**, 4545 (1979).
52. R. Prozorov, R. W. Giannetta, A. Carrington, P. Fournier, R. L. Greene, P. Guptasarma and D. G. Hinks, A. R. Banks, Measurements of the absolute value of the penetration depth in high-$T_c$ superconductors using a low-$T_c$ superconductive coating, *Appl. Phys. Lett.* **77**, 4202 (2000).
53. N. E. Phillips and R. A. Fischer, Superconducting-state energy gap parameters from specific heat measurements, *J. Therm. Anal. Calorim.* **81,** 631 (2005).
54. D. F. Moore, R.B. Zubeck, J.M. Rowell and M.R. Beasley, Energy gaps of the A-15 Superconductors $Nb_3Sn$, $V_3Si$, and $Nb_3Ge$ measured by tunneling, *Phys. Rev. B* **20**, 2721 (1979).